\documentclass[12pt,manuscript,authoryear]{aastex}

\usepackage{graphicx}
\usepackage{natbib}
\usepackage{epstopdf}
\newcommand{\app}{\ensuremath{\sim} }

\begin{document}

\title{The effect of multiple particle sizes on cooling rates of chondrules produced in large-scale shocks in the solar nebula}
  
\author{{\it Short Title: Chondrule Cooling Rates ~~~~Article Type: Journal}}

\author{Melissa~A.~Morris}
\affil{State University of New York at Cortland,
        P.~O.~Box 2000, Cortland, NY 13045-0900}
\affil{School of Earth and Space Exploration, Arizona State University,
        P.~O.~Box 876004, Tempe, AZ 85287-6004}
\email{melissa.morris@cortland.edu}

\and

\author{Stuart J. Weidenschilling}
\affil{Planetary Science Institute,
        1700 East Fort Lowell, Suite 106, Tucson, AZ 85719-2395 }

\and

\author{Steven J. Desch}
\affil{School of Earth and Space Exploration, Arizona State University,
        P.~O.~Box 871404, Tempe, AZ 85287-1404}

\begin{abstract} 
Chondrules represent one of the best probes of the physical conditions and processes acting in the early solar nebula.  Proposed chondrule formation models are assessed based on their ability to match the meteoritic evidence, especially experimental constraints on their thermal histories.  The model most consistent with chondrule thermal histories is passage through shock waves in the solar nebula.  Existing models of heating by shocks generally yield a good first-order 
approximation to inferred chondrule cooling rates. However, they predict 
prolonged heating in the pre-shock region, which would cause volatile loss 
and isotopic fractionation, which are not observed. These models have
typically included particles of a single (large) size, i.e., chondrule
precursors, or at most, large particles accompanied by micron-sized grains. The size distribution of solids present during chondrule formation controls the opacity of the affected region, and
significantly affects the thermal histories of chondrules. Micron-sized
grains evaporate too quickly to prevent excessive heating of chondrule
precursors. However, isolated grains in chondrule-forming 
regions would rapidly coagulate into fractal aggregates. Pre-shock 
heating by infrared radiation from the shock front would cause these aggregates
to melt and collapse into intermediate-sized (tens of microns) particles. We show that inclusion of such particles yields chondrule cooling rates consistent with petrologic and isotopic constraints.        
\end{abstract}


\label{firstpage}

\section{Introduction}

Chondrules are submillimeter- to millimeter-sized igneous spherules found in chondritic meteorites.  They formed as free-floating objects in the solar nebula, prior to their subsequent incorporation into chondrite parent bodies (Lauretta et al. 2006).  The silicate chondrule precursors, either single particles or aggregates of smaller particles, 
were ``flash heated" to the melting point and then cooled through the crystallization point.
Later, these objects were incorporated into chondrites through a process apparently involving 
aerodynamic sorting (Teitler et al.\ 2010).  Isotopic dating of chondrules places their formation over a wide range in time; from contemporaneous with the formation of the oldest solids, the calcium-aluminum-rich inclusions (CAIs), to as long as 6 Myr after CAIs (Amelin et al. 2002; Kita et al. 2005; Krot et al. 2005; Russell et al. 2006; Wadhwa et al. 2007; Connelly et al. 2008; Krot et al. 2008; Amelin et al. 2010; Connelly et al. 2012; Bollard et al. 2013; Bollard et al. 2014).  The formation of most chondrules falls within a period 2-3 Myr after CAIs (Kurahashi et al. 2008; Villeneuve et al. 2009), comparable to the lifetimes of protoplanetary disks (3-6 Myr; Haisch et al. 2001; Williams \& Cieza 2011).  Because the great majority of chondrules formed at the dawn of the Solar System, they are ideal probes into the physical conditions and processes acting in the early solar nebula.

Several reviews of chondrules and the chondrule formation process have been presented (Jones et al.\ 2000;
Connolly et al.\ 2006; Lauretta et al.\ 2006; Desch et al.\ 2012), and 
numerous models have been proposed for chondrule formation, with nebular shocks considered the leading mechanism (e.g. Iida et al. 2001; Ciesla \& Hood 2002; Desch \& Connolly 2002; Morris \& Desch 2010, Morris et al. 2012).  A substantial amount of information in the meteoritic record serves to constrain the formation mechanism.  Meteoritic constraints on chondrule formation were divided by Desch et al.\ (2012) into two broad 
categories:  those pertaining to chondrule thermal histories, and ``non-thermal" constraints. Models are judged by their ability to meet these constraints, especially the inferred thermal histories of chondrules.     

The mechanism found to be most consistent with the constraints is passage of chondrule precursors through nebular shocks (Desch et al. 2005; Desch et al, 2010; Desch et al. 2012).  Large-scale shocks, such as those driven by gravitational instabilities (e.g., Boss \& Durisen 2005; Boley \& Durisen 2008), are particularly consistent with the meteoritic constraints, both non-thermal and thermal, with the exception of excessive heating in the pre-shock region (Morris \& Desch 2010).   
%Bow shocks around planetary embryos are compatible with chondrule formation (Morris et al. 2012), as well as large-scale nebular shocks, such as those driven by gravitational instabilities (e.g., Boss \& Durisen 2005; Boley \& Durisen 2008).  Both shock mechanisms generally conform to the constraints on the thermal histories of chondrules, with the exception of excessive heating in the pre-shock region in the case of large-scale nebular shocks (Morris \& Desch 2010).  
%The long pre-shock heating stage in a large-scale shock is one of the main differences between large-scale shocks and bow shocks around planetary embryos.  
This pre-heating occurs because a radiation front, known as a Marshak wave, penetrates into the pre-shock region from the hot, post-shock region, and is well-described by treating the radiation field in the diffusion approximation (Mihalas \& Mihalas 1984). The distance the Marshak wave penetrates is determined by how quickly the infrared radiation can diffuse, which is, in turn, determined by the opacity of the material.  Using standard nebular opacity and only one, large precursor size (characteristic of chondrules) as in Morris \& Desch (2010), it is inevitable that chondrule precursors will be heated for hours before they are overrun by the shock front.  If so, they would lose volatiles like sulfur and sodium, and experience Rayleigh fractionation, which are not observed in the meteoritic record.  Higher opacity in the pre-shock region would halt the Marshak wave closer to the shock front and eliminate excess pre-heating of chondrule precursors.

Opacity in the pre-shock region is determined by the size and abundance of particles present.  Models to-date have, at most, included micron-sized dust and chondrule precursors of a single large (300 $\mu$m) size (Iida et al. 2001; Ciesla \& Hood 2002; Desch \& Connolly 2002; Morris \& Desch 2010).  Higher concentrations of large precursors that survive the passage of the shock would increase post-shock cooling rates of chondrules to the point that they no longer meet the meteoritic constraints (Desch \& Connolly 2002; Morris \& Desch 2010).  Alternatively, micron-sized dust is expected to vaporize too readily to inhibit the pre-shock heating (Morris \& Desch 2010).  Here we explore one 
possibility that may shorten the duration of pre-shock heating, by proposing that a 
third population of intermediate-sized (\app10 $\mu$m) particles exists in the pre-shock 
region. Based on our understanding of dust coagulation in the solar nebula (Sect. 5) 
we expect micron-sized grains to form fractal aggregates in the pre-shock region.  We assume that once the particles start to melt, surface tension will pull them into compact spheres, prolonging their lifetimes against evaporation.  Although this assumption is supported by the work of Ebel (2006), experimental verification is warranted.  These intermediate-sized particles would add significant opacity to the pre-shock region, reducing the duration of pre-heating.
 We also anticipate that these intermediate-sized particles would largely evaporate at or near the shock front, thereby having little effect on the post-shock cooling rates of chondrules.     

In this paper we include intermediate-sized particles in the model of Morris \& Desch (2010), and present our predictions of the effects on chondrule thermal histories.  In comparing our results with the meteoritic record, we conclude that a reasonable abundance of intermediate-sized particles in the chondrule-forming region will eliminate the otherwise hours-long pre-shock heating, with little effect on post-shock thermal histories.  

\section{Constraints on Chondrule Formation}

%\subsection{"Non-Thermal" Constraints on Chondrule Formation}

Among the non-thermal constraints on chondrule formation are the following:
The retention and lack of isotopic fractionation of volatiles imply the size of the chondrule-forming 
region was $\gg 10^3$ km and the density of chondrules exceeded $\sim 10 \, {\rm m}^{-3}$ 
(Cuzzi \& Alexander 2006). 
The observed frequency of compound chondrules likewise argues for a chondrule-forming region 
$\gg 10^3$ km and chondrule density $> \, 10 \, {\rm m}^{-3}$ (Desch et al.\ 2012).
High pressures $> 10^{-3}$ atm are inferred, to stabilize the silicate melt against too-rapid evaporation
(Ebel \& Grossman 2000).
The oxygen fugacity in the chondrule-forming region is thought to have been variable, but was likely more oxidizing 
than a solar composition gas (Krot et al.\ 2000). 
As outlined above, most chondrules formed 2-3 Myr after CAIs, spanning an even greater time range.
Chondrule formation was not a one-off event but an ongoing process, as evidenced by fragments of relict 
chondrules within later generations of chondrules (Jones et al.\ 2000).
The chemical complementarity of chondrules and matrix has been interpreted as meaning that chondrules and 
matrix dust within a given chondrite formed in the same region of the solar nebula, with both sampling from a 
solar-composition reservoir (Wood 1985; Palme et al.\ 1993; Klerner \& Palme 1999; Hezel \& Palme 2008).
However, it is very difficult to determine whether these complementary reservoirs were produced in the solar
nebula or by redistribution on the parent body (Zanda et al.\ 2009), and this constraint remains controversial. 

%\subsection{Constraints on Thermal Histories of Chondrules}

The chemistry and textures of chondrules constrain their thermal histories.  Peak temperature is determined by assessing the number of nuclei remaining in the melt and/or the number of nuclei encountered as external seed nuclei (Lofgren 1982, 1989, 1996; Hewins \& Connolly 1996; Hewins 1997; Connolly et al.\ 1998; Desch \& Connolly 2002; Lauretta et al.\ 2006).  Cooling rates are constrained by texture, major and minor element abundances, and bulk chemistry (Connolly \& Desch 2004; Connolly et al.\ 2006; Lauretta et al.\ 2006).   The majority of chondrules (porphyritic texture; 84\% of ordinary chondrite chondrules; Gooding \& Keil 1981), experienced peak temperatures in the range of 1770 - 2120 K for several seconds to minutes (Lofgren \& Lanier 1990; Radomsky \& Hewins 1990; Hewins \& Connolly 1996; Lofgren 1996; Hewins 1997; Connolly \& Love 1998;  Jones et al.\ 2000; Connolly \& Desch 2004; Hewins et al.\ 2005; Ciesla 2005; Connolly et al.\ 2006; Lauretta et al.\ 2006).  Peak temperatures for barred textures ranged between 1820 - 2370 K (Hewins \& Connolly 1996). According to furnace experiments, porphyritic chondrules cooled at rates between $10-1000 \, {\rm K} \, {\rm hr}^{-1}$ (Hewins et al.\ 2005), and $\sim 300 - 3000 \, {\rm K} \, {\rm hr}^{-1}$ for barred textures (see review by Desch et al.\ 2012).  The prevalence of barred textures among compound chondrules argues for higher cooling rates in regions of high chondrule density (Desch \& Connolly 2002; Desch et al.\ 2012).
Chondrules retain volatile elements, such as Na and S, indicating that they did not remain above the liquidus for more than minutes, and cooled quite rapidly ($\gtrsim 10^4 \, {\rm K} \, {\rm hr}^{-1}$) to their liquidus temperature (Yu \& Hewins 1998).  The presence of primary S tells us that chondrules formed in an environment with an ambient temperature $<$ 650 K (Rubin et al. 1999) and did not experience prolonged heating between \app 650 - 1200 K for more than several minutes (Hewins et al.\ 1996; Connolly \& Love 1998; Jones et al.\ 2000; Lauretta et al.\ 2001; Tachibana \& Huss 2005; Connolly et al.\ 2006).  Additionally, there is no indication of the isotopic fractionation that would arise from the free evaporation of alkalis such as Na, which constrains the time spent at high temperature before melting (Tachibana et al.\ 2004).  Modeling of isotopic fractionation has demonstrated that chondrules were heated from 1300 to 1600 K in times on the order of minutes, in order to prevent isotopic fractionation of S (Tachibana \& Huss 2005).  It is important to note that such constraints on volatile loss and fractionation may be negated if the partial pressures of volatiles are high in the chondrule-forming environment, allowing suppression of evaporation or recondensation.  However, in the absence of an environment that is considerably enriched in volatile or moderately volatile elements (Connolly et al.\ 2006), the time constraint on ``pre-heating" of chondrules must be met.  %For the remainder of this paper, we assume this constraint to be valid. 

Interpreting these petrographic and other data about chondrules requires the context of a physical model 
for their formation.
Numerous models of chondrule formation have been proposed (see, for example, the review by Desch et al.\ 2012).
As emphasized by Connolly \& Desch (2004) and by Desch et al.\ (2012), a successful model must make 
quantitative predictions about chondrule properties that can be tested against the substantial meteoritic data.
As discussed in Desch et al.\ (2012), ``non-thermal" constraints are somewhat diagnostic, ruling out some models, 
but only weakly so. 
Chondrule formation models such as melting by lightning, the X-wind model, and shocks due to X-ray flares or accretion, have difficulty meeting one or more of the non-thermal constraints  (e.g. size of the region, chronology, gas density; Desch et al. 2012). 
%Models in which chondrule formation is due to melting by lightning in the solar nebula are consistent
%with the chronology of chondrule formation and with chondrule-matrix complementarity, but tend to predict 
%small formation regions and low pressures, although only the size of the region is a firm constraint. 
%The X-wind model of chondrule formation as presented is inconsistent with the chronology of chondrule
%formation (see also Desch et al.\ 2010), and definitely predicts non-complementarity of chondrules 
%and matrix, although parent-body redistribution may nullify this constraint.
%Some shock models of chondrule formation can also be excluded, such as those that rely on chondrule
%precursors being melted in the low-density gas at the disk surfaces, e.g., by shocks driven by X-ray 
%flares (Nakamoto et al.\ 2005), clumpy accretion onto the top of the disk (Boss \& Graham 1993),
%or an accretion shock at the top of disk (Ruzmaikina \& Ip 1994).
%Likewise, accretion shocks in the Jovian subnebula, proposed by Nelson \& Ruffert (2005), also appear
%to involve gas too low in density (Desch et al.\ 2012).
In contrast, shocks at the disk midplane, driven by gravitational instabilities (spiral density waves) in 
the nebula (Wood 1996, 1996; Desch \& Connolly 2002; Boss \& Durisen 2005; Boley \& Durisen 2008), or bow shocks in front of planetary bodies on eccentric orbits (Hood 1998; Weidenschilling et al.\ 1998; 
Ciesla et al.\ 2004; Hood et al.\ 2005, 2009; Hood \& Weidenschilling 2012; Morris et al.\ 2012; Boley et al. 2013), remain consistent with most non-thermal constraints.  

Constraints on the thermal histories of chondrules are more diagnostic and can be used to definitively 
rule out some models.
The X-wind model predicts ambient temperatures that are too high ($\approx 900 \, {\rm K}$), peak 
temperatures that are too low ($\ll 2000 \, {\rm K}$), and a single, low cooling rate 
$< 10 \, {\rm K} \, {\rm hr}^{-1}$ for all chondrules at all temperatures below the peak (inconsistent
with barred olivine chondrules), and no correlation of cooling rate with chondrule density
(Desch et al.\ 2010).
The thermal histories of chondrules are not consistent with melting by nebular lightning,
as chondrules are predicted to cool from the peak all the way to ambient temperatures at rates 
$> 10^4 \, {\rm K} \, {\rm hr}^{-1}$, again with no correlation of cooling rate with chondrule density
(Desch et al.\ 2012).
Detailed thermal histories have not yet been computed for chondrules melted in all types of nebular 
shocks, but in general they would appear to agree with the meteoritic constraints, especially if gas 
densities and chondrule concentrations are as high as implied by the other constraints. 
Models in which chondrules form in the bow shocks around planetesimals (radii of a few hundred km) 
are known to predict chondrule cooling rates through the crystallization temperature range that are 
too high ($\gg 10^3 \, {\rm K} \, {\rm hr}^{-1}$) to yield porphyritic textures (Ciesla et al.\ 2004). 
The 2-D model of Morris et al.\ (2012) demonstrated that thermal histories of chondrules melted in bow shocks
around large (radius $\sim 3000$ km) planetary embryos may be less than 
$10^3 \, {\rm K} \, {\rm hr}^{-1}$.
Recent results from the 3-D model of Boley et al.\ (2013) show that solids melted in bow shocks driven by large planetary bodies
cool at faster rates, due to adiabatic cooling in the rarefaction wave behind the planetary embryo.
Their work shows that solids melted in bow shocks may not even cool in a monotonic fashion.
So although bow shocks around large protoplanets remain a viable formation mechanism for some textures of chondrules, it is not clear that they are responsible for the majority (porphyritic; Boley et al. 2013).

\section{Chondrule Thermal Histories in Large-Scale Shocks}

Large-scale shocks driven by gravitational 
instabilities remain the most promising candidate for the formation of the majority of chondrules. 
These shocks are characterized by lateral lengthscales $\gg 10^6$ km and are therefore largely 1-D in nature.  
This difference in geometry practically eliminates the cooling associated with adiabatic expansion seen
in bow shock models, and also limits the ability of the gas to radiate to a cooler region. 
Cooling rates are set instead mainly by the ability of the post-shock gas to radiate parallel to the 
gas flow, into the pre-shock gas.  
Because of these differences, the thermal histories of chondrules melted in large-scale shocks are the
best match to the meteoritic constraints.

A typical thermal history is shown in Figure 1.
The red line represents the thermal histories suggested by the meteoritic record.  The chondrule precursors start at a low
ambient temperature (set to 600 K in order to meet the meteoritic constraints; Rubin et al. 1999), then heat to their peak temperature in $< 0.3 \, {\rm hr}$,
as constrained by the lack of isotopic fractionation of S (Tachibana \& Huss 2005).
%For an assumed liquidus temperature of 1900 K, within the range of liquidus temperatures for chondrule 
%compositions (Radomsky \& Hewins 1990), a peak temperature $\approx 2000 \, {\rm K}$ is inferred to yield 
%porphyritic textures. 
The melted chondrules cool below the liquidus temperature of $\sim$ 1900 K (consistent with liquidus temperatures for chondrule 
compositions; Radomsky \& Hewins 1990) within minutes, at rates 
$\sim 10^3 - 10^4 \, {\rm K} \, {\rm hr}^{-1}$ (Yu \& Hewins 1998), then take hours to cool through the crystallization temperature range.  Figure 1 depicts cooling rates through the crystallization range $\sim 10 \, {\rm K} \, {\rm hr}^{-1}$, consistent with porphyritic textures.

The physics behind models of chondrule formation in large-scale shocks predicts thermal histories very similar
to those implied by the meteoritic record.  
The black curve in Figure 1 shows the prediction of Morris \& Desch (2010), which assumes an ambient gas with density
$\rho_{\rm g} = 1 \times 10^{-9} \, {\rm g} \, {\rm cm}^{-3}$ and temperature 300 K, a population of micron-sized
dust grains with a mass fraction $1.25 \times 10^{-3}$ relative to the gas, and a population of chondrules
with radii $a_{\rm c} = 300 \, \mu{\rm m}$ and a mass fraction $(3.75 \times 10^{-3}) \, {\cal C}$, where ${\cal C}$  
is the concentration of chondrules, relative to the solar abundance of condensable elements at that temperature.  In this case, ${\cal C} = 10$. 
Through this system, a shock propagates at a speed of $8 \, {\rm km} \, {\rm s}^{-1}$. 
Chondrule precursors start at the temperature of the ambient gas.
As the shock approaches, they are heated by infrared radiation emitted by solids behind the shock.
During this stage, described in more detail below, chondrules are heated nearly to their peak temperatures. 
At the shock front, the gas velocity instantaneously changes (within meters) but the chondrule precursors only 
change velocity after colliding with their own mass of gas.
Therefore, they reach the gas velocity only after the aerodynamic stopping time, 
$t_{\rm stop} \sim (\rho_{\rm s} a) / (\rho_{\rm g} c_{\rm g}) \sim 1$ minute
(here $\rho_{\rm s} \approx 3 \, {\rm g} \, {\rm cm}^{-3}$ is the material density of the chondrules,
$a = 300 \, \mu{\rm m}$ their radii, $\rho_{\rm g} \approx 6 \times 10^{-9} \, {\rm g} \, {\rm cm}^{-3}$
the post-shock gas density, and $c_{\rm g} \approx 3 \, {\rm km} \, {\rm s}^{-1}$, the post-shock sound speed).
For this short duration, chondrules are not only heated by thermal exchange with the shocked gas and absorption
of infrared radiation, but also by frictional heating.  
The frictional heating raises their temperatures by a few $\times 10^2$ K, to their peak temperatures, but
this temperature gain is lost within a time $t_{\rm stop}$ after passing through the shock, implying cooling rates from
the peak $\sim 10^4 \, {\rm K} \, {\rm hr}^{-1}$.
Thereafter, chondrules are in good thermal equilibrium with the gas, and the gas and chondrules cool only as
fast as they can leave the shock front.  
This requires them to move several optical depths downstream.
Because smaller particles evaporate in the pre-shock region, the opacity in the post-shock region is set 
by chondrules only, naturally explaining why the cooling rate of chondrules scale with
the chondrule density (Desch \& Connolly 2002). 

Morris \& Desch (2010) find cooling rates of $10-25 \, {\rm K} \, {\rm hr}^{-1}$ through the crystallization temperature range, for chondrule concentrations ${\cal C} \sim 10$.  These cooling rates are consistent with those needed to produce poprphyritic 
textures.   For ${\cal C} \sim 10^2 - 10^3$, the cooling rates are consistent with barred olivine textures.
This may explain why barred textures are much more common (relative to porphyritic textures) among compound 
chondrules, which will preferentially form in regions with high ${\cal C}$.

\section{Opacity and Evaporation}

The agreement between the predicted thermal history shown in Figure 1 and the thermal histories inferred from meteoritic
constraints is quite good, with the exception of the duration of heating of chondrule precursors 
in advance of the shock.
As is evident from Figure 1, for typical parameters (e.g., Morris \& Desch 2010), this pre-shock heating
stage lasts roughly 10 hours.   Precursors are above the solidus temperature ($\approx 1400 \, {\rm K}$) for
approximately 4 hours. 
That chondrules will be heated before the shock is a robust feature of these
1-D shock models, and is the inevitable consequence of infrared radiation diffusing into the pre-shock region. 
The diffusion of radiation ``upstream", into the pre-shock region, is a well understood phenomenon in 
radiation hydrodynamics called a Marshak wave (Mihalas \& Mihalas 1984).
% FIGURE 1
\begin{figure}[ht]
\includegraphics[width=14cm]{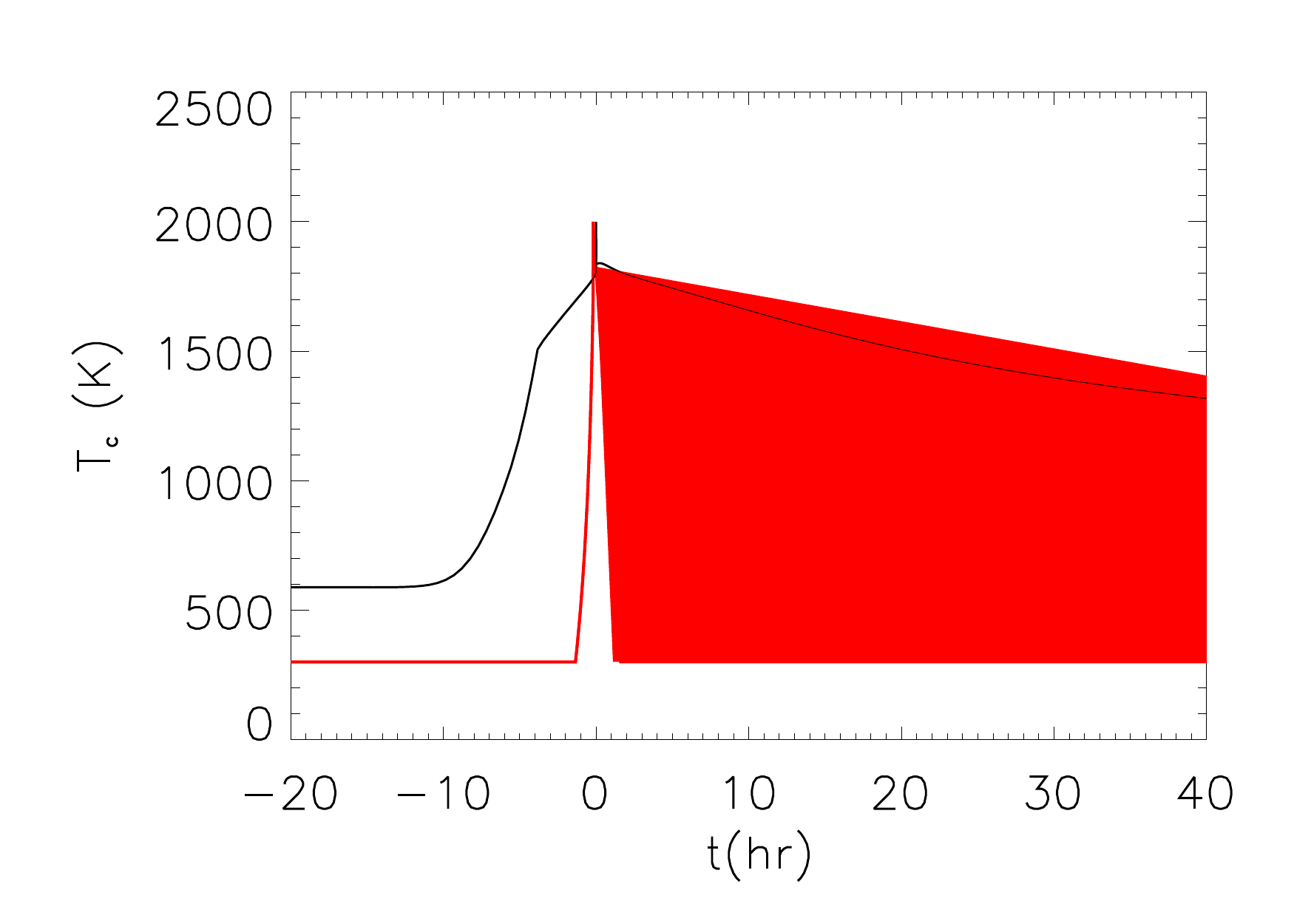}
\caption{The black curve indicates chondrule thermal histories as predicted by the shock model of Morris \& Desch (2010), as compared to those inferred from experimental constraints (red curve).  The shaded region in the post-shock region indicates the range of inferred cooling rates (10-1000 K/hr).  Chondrules start at an ambient temperature of 300K, but have already reached \app 600K ten hours before reaching the shock front. }
\label{fig:history1}
\end{figure} 
%It can be seen from the figure that chondrules begin heating up from the ambient temperature of 300 K more than 10 hours before they reach the shock front, at an initial rate of about $250 \, {\rm K} \, {\rm hr}^{-1}$.
%At 3.8 hours before reaching the shock front, the chondrules have already reached 1507 K, and heat up to 1805 K immediately 
%before entering the shock.
The distance $R$ between the shock front and the edge of this radiation front is quantified by 
comparing the rate of propagation by radiative diffusion, $d R / dt = ({\cal D} / t)^{1/2} / 2$, with the
rate at which cold pre-shock gas is advected into the flow, $d R / dt = V_{\rm s}$, giving $({\cal D}/t)^{1/2} = 2V_{\rm s}$.  
Here ${\cal D}$ is the radiation diffusion coefficient pertinent to a dusty medium, defined below.
From this we see that the radiation front will stop when $t$, the time for material to flow from the radiation
front to the shock front, is equal to $t \sim {\cal D} / 4 V_{\rm s}^{2}$.
Therefore, the time chondrule precursors spend in the heated pre-shock region (also $t$) depends not only on the shock speed, but the radiation diffusion coefficient as well. 

The radiation diffusion coefficient in a gaseous medium with density $\rho_{\rm g}$, temperature $T$, and 
heat capacity $C_{\rm g}$, is
\begin{equation}
{\cal D} = \frac{ 64 \pi^2 \, \lambda \, \sigma_{\rm SB} T^4 }{ 3 (\rho_{\rm g} C_{\rm g} T) },
\end{equation} 
where $\sigma_{\rm SB}$ is the Stefan-Boltzmann constant and $\lambda$ is the mean free path of photons in
the medium (Spiegel 1972; Kippenhahn \& Weigert 1990; Desch \& Connolly 2002).
The diffusion coeefficient must also account for the specific heat of solids, $C_{{\rm s}} = 10^{-7}$ erg g$^{-1}$ K$^{-1}$, as well as their spatial density, $\rho_{\rm d} = 10^{-8}$ cm$^{-3}$, giving
\begin{equation}
{\cal D} = \frac{64 \pi^2 \lambda \sigma_{\rm SB} T^3}{3 (\rho_{\rm g} C_{\rm g}+ \rho_{\rm d}C_{\rm s})}
\end{equation}  
This formula takes into account that radiation does more than scatter through the dusty medium.  It is also 
absorbed, warming the gas and solids, which then reemit the radiation.
The mean free path of the photons is equal to $\lambda = 1 / (\rho_{\rm g} \kappa)$, where $\kappa$ is the
opacity of the medium.

Combining these results, 
\begin{equation}
{\cal D} = \frac{ 64 \pi^2 \, \sigma_{\rm SB} T^3 }{ 3 (\rho_{\rm g} \kappa) \, (\rho_{\rm g} C_{\rm g}+ \rho_{\rm d}C_{\rm s}) }.
\end{equation}
Thus ${\cal D}$ and $t$ (the duration of the pre-shock heating) are inversely proportional to the square of 
the density, and also inversely proportional to the opacity.  Therefore, higher opacity will result in a shorter duration of pre-shock heating.
%For typical parameters (heat capacity $1.44 \times 10^8 \, {\rm erg} \, {\rm g} \, {\rm K}^{-1}$, average
%pre-shock temperature $T = 1400 \, {\rm K}$), and assuming $\kappa = 4.5$ (see below), 

%\begin{equation}
%t = 3.23 \, 
% \left( \frac{ \rho_{\rm g} }{ 10^{-9} \, {\rm g} \, {\rm cm}^{-3} } \right)^{-2} \,
% \left( \frac{ \kappa }{ 4.5 \, {\rm cm}^{2} \, {\rm g}^{-1} } \right)^{-1} 
% \left( \frac{ V_{\rm s} }{ 8 \, {\rm km} \, {\rm s}^{-1} } \right)^2 \, 
% \, {\rm hr}.
%\end{equation}
 
%\noindent This is in good accord with the behavior exhibited in Figure 1. 

The pre-shock gas density and shock speed have a large bearing on the peak heating experienced by the 
chondrules, and are relatively well-constrained.
The major uncertainty in setting the timescale for pre-shock heating is the opacity. 
If the opacity in the pre-shock region is attributable to spherical particles with radius $a$,
mass $m = (4\pi/3) \rho_{\rm s} a^3$, and total spatial density $\rho_{\rm d}$, then the opacity is given by 
\begin{equation}
\kappa = \frac{ (\rho_{\rm d} / m) \, \pi a^2 \, Q_{\rm abs}}{ \rho_{\rm g} } 
 = \frac{3 \, (\rho_{\rm d} / \rho_{\rm g}) \, Q_{\rm abs}}{4\, \rho_{\rm s} a},
\end{equation}
where $Q_{\rm abs}$ is the wavelength-averaged absorption efficiency, which is well approximated by 0.8 for
wavelengths $\lambda < 2\pi a$, as is the case for radiation inside the Marshak wave.
For micron-sized dust ($a = 0.5 \, \mu{\rm m}$) made of silicates ($\rho_{\rm s} = 3.3 \, {\rm g} \, {\rm cm}^{-3}$),
with a dust-to-gas mass ratio $1.25 \times 10^{-3}$ (as the model shown in Figure 1 assumed), 
$\kappa = 4.5 \, {\rm cm}^2 \, {\rm g}^{-1}$.
The opacity of chondrules ($a_{\rm c} = 300 \, \mu{\rm m}$) with mass ratio 
$\rho_{\rm c} / \rho_{\rm g} = 3.75 \times 10^{-3} \, {\cal C}$ is 
$\kappa = 0.2 \, ({\cal C} / 10) \, {\rm cm}^{2} \, {\rm g}^{-1}$, meaning that dust grains dominate the 
opacity.
%Assuming a dust-to-gas ratio $\rho_{\rm d} / \rho_{\rm g} = 5 \times 10^{-3}$, a particle radius 
%$a_p = 0.5 \, \mu{\rm m}$, and an absorptivity $Q_{\rm abs} = 1$ for $\lambda < 2\pi a_s$ and 
%$Q_{\rm abs} = 2\pi a_s/\lambda$ for $\lambda > 2\pi a_s$, gives opacity due to dust of
%\begin{equation}
%\kappa_{\lambda} = 30 \, \min \left[ 1, (\lambda / 3.1 \, \mu{\rm m})^{-1} \right] \, {\rm cm}^{2} \, {\rm g}^{-1}, 
%\label{eq:ourkappa1}
%\end{equation}
%Morris \& Desch (2010) derived an approximation to the wavelength-averaged opacity over the temperature range relevant to chondrule formation.  For a solar composition ($\rho_{\rm d} / \rho_{\rm g} = 5 \times 10^{-3}$), they found 
%\begin{equation} 
%\kappa_{\rm{app}}= 12.161 \; \ln(T / 1 \, {\rm K}) - 62.524 \; {\rm cm}^{2} \; {\rm g}^{-1}\, 
%\label{appkappa}
%\end{equation} 
%per gram of gas.
%Alternatively, the opacity due to large chondrule precursors is only ($\kappa$ = 0.03 cm$^2$ g$^{-1}$) (Morris \& Desch 2010), meaning that dust grains dominate the opacity in the pre-shock region.  
However, micron-sized dust is predicted to 
vaporize when the temperature reaches about 1500 K (Morris \& Desch 2010).  This is why the temperature gradient becomes less steep about 4 hours before the fluid encounters the shock front, as shown in Figure 1.  This early evaporation of dust and resulting drop in opacity (with only large, chondrule-sized particles as the source) is responsible for the prolonged pre-heating of chondrule precursors.    

These features of the 1-D shock are quite robust.  
Radiation must propagate into the pre-shock region as a Marshak wave.
The time fluid spends in this heated pre-shock region is $t = {\cal D} / (4 V_{\rm s}^{2})$.
Radiation essentially propagates into the pre-shock region a number of optical depths
$\tau = \rho_{\rm g} \kappa V_{\rm s} t \sim 10^2$.
For chondrules to reach peak temperatures $\approx 2000 \, {\rm K}$ after the shock, the temperature must
be $\approx 1800 \, {\rm K}$ before the shock hits, and therefore micron-sized dust grains must evaporate
before the shock (Morris \& Desch 2010). 
The opacity due to chondrule precursors alone is small enough that $t \propto \kappa^{-1}$ must be 
at least several hours. 
This result is difficult to reconcile with the short ($< 30$ minutes) pre-shock heating durations
implied by the lack of isotopic fractionation of S (Tachibana \& Huss 2005).

%The Marshak wave propagates a distance, $R$, into the pre-shock region from the hot, post-shock region at the same rate that new material moves into it.  The rate that the Marshak wave diffuses into the pre-shock region is given by $dR/dt = 1/2 \sqrt{D/t}$, where $D$ is the diffusion coefficient, and $t$ is the time.  The diffusion coefficient, 
%\begin{equation}
%D = \frac{64 \pi^2 \lambda \sigma T^3}{3 \rho_{\rm s}C_{V_s}C_{V_g}}= \frac{64 \pi^2 \sigma T^3}{3 \rho_{\rm s}\rho_{\rm g}\kappa C_{V_s}C_{V_g}},
%\end{equation} 
%\noindent where $\lambda$ is the mean free path of a photon, $\sigma$ is the Stefan-Boltzmann constant, $T$ is the temperature, $\rho_{\rm s}$ is the material density of solids, $\rho_{\rm g}$ is the gas density, $\kappa$ is the opacity, $C_{V_{\rm s}}$ is the heat capacity of solids, and $C_{V_{\rm g}}$ is the heat capacity of the gas.  The Marshak wave is halted when $dR/dt = V_{\rm s}$, giving $\sqrt{D/t}= 2V_{\rm s}$, so that $t= D/2V_{\rm s}$ and $R = 2V_{\rm s}t = (1/4)D/2V_{\rm s}$.  For $\kappa$ = 0.3 cm$^2$ g$^{-1}$, $R$ \app 375,000 km and $t$ \app 13 hr; for $\kappa$ = 1.0 cm$^2$ g$^{-1}$, $R$ \app 113,000 km and $t$ \app 4 hr, comparable to the results of Morris \& Desch (2010).  

Due to the opacity dependence of the diffusion coefficient, higher opacity in the pre-shock region would halt the Marshak wave closer to the shock and would reduce the time particles spend at high temperatures.  Increasing the opacity by about a factor of 10 would eliminate the excess pre-heating of chondrule precursors.  In current models (e.g. Morris \& Desch 2010), the only sources of opacity in the pre-shock region are dust and the precursors themselves.  Dust will evaporate well before it reaches the shock front, resulting in a pronounced drop in opacity.  Simply increasing the abundance of large chondrule precursors to the degree necessary to maintain high opacity would assume concentrations 100 times the background dust-to-gas ratio.  These precursors would then increase the opacity in the post-shock region, driving cooling rates to well over 1000 K hr$^{-1}$; rates that are inconsistent with chondrule formation.  However, based on dust coagulation models (see Sect. 5), we expect micron-sized dust to form fractal aggregates consisting of 
thousands of monomers. If a substantial fraction of such aggregates was in a size 
range that provided 
increased opacity in the pre-shock region by surviving the Marshak wave, yet evaporated at the shock front, the problems of excess pre-heating or too-rapid cooling rates would be alleviated.    

In order to find the optimal particle size range necessary, we use the Hertz-Knudson equation to determine the temperature-dependent evaporation rate for small and intermediate-sized particles: 
\begin{equation} 
J_i= \sum_{j=1}^n \; \frac{n_{ij}\gamma_{ij}P_{ij}^{sat}}{\sqrt{2 \pi m_{ij}RT}},
\label{eq:evap}
\end{equation} 
in units of mol cm$^{-2}$ s$^{-1}$, where $i$ is the isotope or element considered, $j$ is the gas species 
containing $i$, $n$ is the number density of $i$, $\gamma$ is the evaporation coefficient of $i$, $P^{sat}$ is 
the saturation 
vapor pressure for $j$, $m$ is the molecular weight of $j$, $R$ is the gas constant, and $T$ is the temperature.
Davis \& Richter (2005) give the temperature-dependent evaporation coefficients for forsterite, and calculate 
the vacuum evaporation rate as a function of temperature, as well as the evaporation rate at 1773 K as a function 
of pressure.
We neglect the temperature dependence of the saturation vapor pressure (a small effect, at most a few times 
$10^{-2}$ g cm$^{-2}$ s$^{-1}$), and substitute the appropriate ratio $\gamma/T$ into the given evaporation rate 
at 1773 K to calculate the evaporation rates at other temperatures.  
At 10$^{-3}$ bar, we find a forsterite dust grain with radius $a = 0.5 \mu$m will evaporate in $<$ 60 s upon 
reaching a temperature $T \, \approx \, 1500$ K.  
By the time the dust has achieved a temperature of \app 1800 K, it will evaporate in $<$ 10 s.
Alternatively, for a particle of radius = 10$\mu$m, it will take $>$ 16 minutes to evaporate at \app 1500 K and \app 3 minutes at \app 1800 K.  

Our calculations show that particles with radii $\sim$ 10, 15, 20, and 30 $\mu{\rm m}$, will likely survive until they reach the shock front, then largely evaporate, consistent with the results of Miura \& Nakamoto (2005).
If so, the cooling rates of chondrules through their crystallization temperature ranges would be unaffected 
and would still be proportional to chondrule concentration ${\cal C}$.  However, the opacity in the pre-shock region 
would be increased significantly.  
%A population of intermediate-sized particles $10 \, \mu{\rm m}$ in radius, with total mass fraction comparable to that
%of dust, $\rho_{\rm d} / \rho_{\rm g} = 1.25 \, \rm x \, 10^{-3}$, would yield an opacity 
%$\kappa = 2.27 \, {\rm cm}^{2} \, {\rm g}^{-1}$, exceeding the opacity due to chondrules, 
%$0.20 \, {\rm cm}^2 \, {\rm g}^{-1}$, by a factor of 10.
%According to our calculations above, the addition of this population of microchondrules would decrease the time chondrule precursors spent above 1400 K in the pre-shock region, from 4 hours to 2 hours.  In reality, the reduction in pre-heating is more dramatic, due to the wavelength-dependency of the opacity.

\section{Particle Sizes in the Meteoritic Record}

Chondrules from all chondrite groups (except CH/CB/Isheyevo) have mean diameters of 150 $\mu$m to 1 mm (Jones et al. 2000 and references therein; Campbell et al. 2005), with characteristic mean diameters for individual chondrite classes following a lognormal size-frequency distribution (Jones et al. 2000; Cuzzi et al. 2010).  Friedrich et al. (2015) and McCain et al. (2015) found broad size distributions, with average particle sizes of $>$ 200 $\mu$m and $>$ 175 $\mu$m respectively. Microchondrules, defined as having diameters $<$ 40 $\mu$m (Krot \& Rubin 1996), have been observed in some instances, including the recent studies of Friedrich et al. (2015) and McCain et al. (2015).  Chondrules in CH chondrites have mean diameters of 20 $\mu$m (Jones et al. 2000).  However, chondrules from CH/CB/Isheyevo-type chondrites are believed to have formed quite differently from the majority of chondrules (e.g. Krot et al. 2005), are mainly skeletal and cryptocrystaline in texture, therefore are not applicable to this study.  Microchondrule fragments were found in samples returned from Comet Wild 2 (Nakamura et al. 2008; Zolensky et al. 2008), and have been observed in unequilibrated clasts and embedded in the fine-grained rims of ``normal-sized" chondrules (Rubin et al. 1982; Krot \& Rubin 1996; Krot et al. 1997; Weisberg \& Ebel 2010; Bigolski et al. 2012; Bigolski et al. 2013; Bigolski et al. 2016).   
%Microchondrules found in the fine-grained rims of NWA 5717, Semarkona, and Bishunpur show cryptocrystalline, microporphyritic, and glassy textures (Bigolski et al. 2013).  
Bigolski et al. (2013; 2016) found that $\sim$ 60\% of all chondrules in NWA 5717 have fine-grained rims, and $\sim$ 20\% of these contained microchondrules $<$ 20 $\mu$m in size (Bigolski et al. 2012).  Yet chondrules of such small size are rarely found in the matrix of chondrites.  The highest abundance reported to-date is $\sim$ 0.08 vol\%, found in the fine-grained matrices of two unequilibrated ordinary chondrites (Dobric\u{a} \& Brearley 2013).  Bigolski et al. (2013; 2016) suggest that microchondrules found in fine-grained rims were formed either by ablation of silicate liquid from the surface of the host chondrule, or by melting of dust nearby.  The spherical morphology and presence of glassy material lead Dobric\u{a} \& Brearley (2013) to the conclusion that microchondrules found in low abundance in the matrix of UOCs must have formed as free-floating objects in the nebula.  Broad size distributions of chondrules would be expected from coagulation models (see below), but aside from the observation of the odd microchondrule, the overwhelming majority of solids in the meteoritic record are either large (chondrule-sized) particles or micron-sized dust.   We assume that most particles $<$ 150 $\mu$m are lost from the meteoritic record, either to evaporation or to size-sorting.

Turbulent coagulation models support a fairly broad size distribution, with rapid coagulation in the 1-100 $\mu$m size range, with mm-size aggregates common (Zsom et. al. 2011; Ormel \& Okuzumi 2013; Birnstiel et al. 2014).  Thermal coagulation alone (without turbulence) rapidly produces aggregates up to 500 $\mu$m (Figure 2).  Most models of chondrule-forming environments assume that small
particles were present as micron-sized dust grains, but they were unlikely to
exist as individual particles. The mean collision time between grains of 
diameter $d$ is $t_{\rm coll} \sim 1/(N d^{2 }\rm V)$, where $N$ is the number of grains per 
unit volume and $V$ their mean velocity. In collisions driven by 
thermal motions, $V \sim \left (8kT/\pi m\right)^{1/2}$, where $k$ is the Stefan-Boltzmann constant,
$T$ the local temperature, and $m$ the particle mass. For the parameters discussed
in Sect. 3, the mass fraction 1.25 x 10$^{-3}$ relative to gas with density 10$^{-9}$ g cm$^{-3}$
gives $N$ of order 1 grain cm$^{-3}; d = 10^{-4}$ cm and $T$=300 K implies $t_{\rm coll} \sim$ 1 year.  
Grains of this size readily stick due to van der Waals surface forces,
forming larger aggregates. Thermally driven coagulation causes hierarchical 
cluster-cluster aggregation, yielding fractal aggregates with very low densities 
that decrease with increasing size (Dominik \& Tielens 1997; Blum \& Wurm 2000;
Blum et al. 2000). These extended cobweb-like assemblages have 
optical properties similar to isolated grains. Thus, they would still dominate
the opacity, which would diminish very slowly due to coagulation. 
Thermally driven coagulation proceeds rapidly for the smallest particles, which have the largest
velocities, and slows for larger ones. This yields a characteristic size
distribution with a rather narrow single peak; the size of the peak increases
at a rate that diminishes with time. This thermal size distribution differs from
ensembles of particles that experience destructive collisions, which yield
power-law distributions (Mathis et al. 1977; Weidenschilling 1997). 
  
The presence of fractal grain aggregates has some interesting implications for
the thermal evolution of chondrules in the context of the shock model. Such
aggregates would be heated by the infrared radiation of the approaching shock,
before they could be disrupted by the acceleration at the shock front.
As discussed above, isolated micron-sized grains would evaporate on timescales of
seconds. However, heating of a low-density fractal aggregate to incipient melting 
would allow it to collapse to a compact object due to surface tension. Such a 
droplet would have a lifetime against evaporation of minutes. They would largely
evaporate at the shock front, effectively erasing them from the meteoritic record,
other than the occasional microchondrule. These solids would contribute substantially
to the opacity in the pre-shock region, eliminating the excess pre-heating found 
in previous models (Morris \& Desch 2010).

In Figure 2 we show size distributions due to thermal coagulation computed for 
our nominal parameters starting with micron-sized monomers at $t$=0.  We do not include the effects of fragmentation due to collisions, which will replenish the micron-sized dust population.  Nor do we include the effects of turbulence.  As such, Figure 2 does not represent the complete size distribution expected to be present in the early solar nebula, but is intended to be illustrative of the growth and inferred presence of fractal aggregates in the indicated size ranges.
We are interested in the earliest history of the solar nebula, when GI-driven shocks would be expected.  At 500 and 5000 years, the fractal aggregates have peaks at sizes \app 40 and 200 $\mu$m, respectively.  Also shown are the equivalent sizes of compact spherical droplets that 
would be produced by melting of these aggregates, with peak sizes at \app 15 and 40 
$\mu$m. The presence of micron-sized grains in sufficient concentration to provide the 
requisite opacity upstream from the shock wave probably guarantees the presence of 
intermediate-sized particles close to the shock itself. 
These results suggest that thermal histories of chondrules could be met by shocks recurring in a given location in the nebula at intervals \app 10$^3$ - 10$^4$ years.
%Thus, it is not necessary to assume a separate source of such particles with an origin distinct from that of the small grains. 

%The source of the initial grains in the simulations is 
%not clear. They could be released by collisions among larger planetesimals, and/or
%condensed from vapor in the aftermath of earlier shock waves. In the latter case,
%repeated passage of shocks through a given region of the nebula at intervals of 
%order hundreds to thousands of years could renew the supply of small grains. 
%Grains that condensed in the wake of chondrule-forming shocks might also be
%incorporated into the resulting chondrites, leading to complementarity of 
%compositions between chondrules and matrix material.

% FIGURE 2
\begin{figure}[t]
\includegraphics[width=8.75cm]{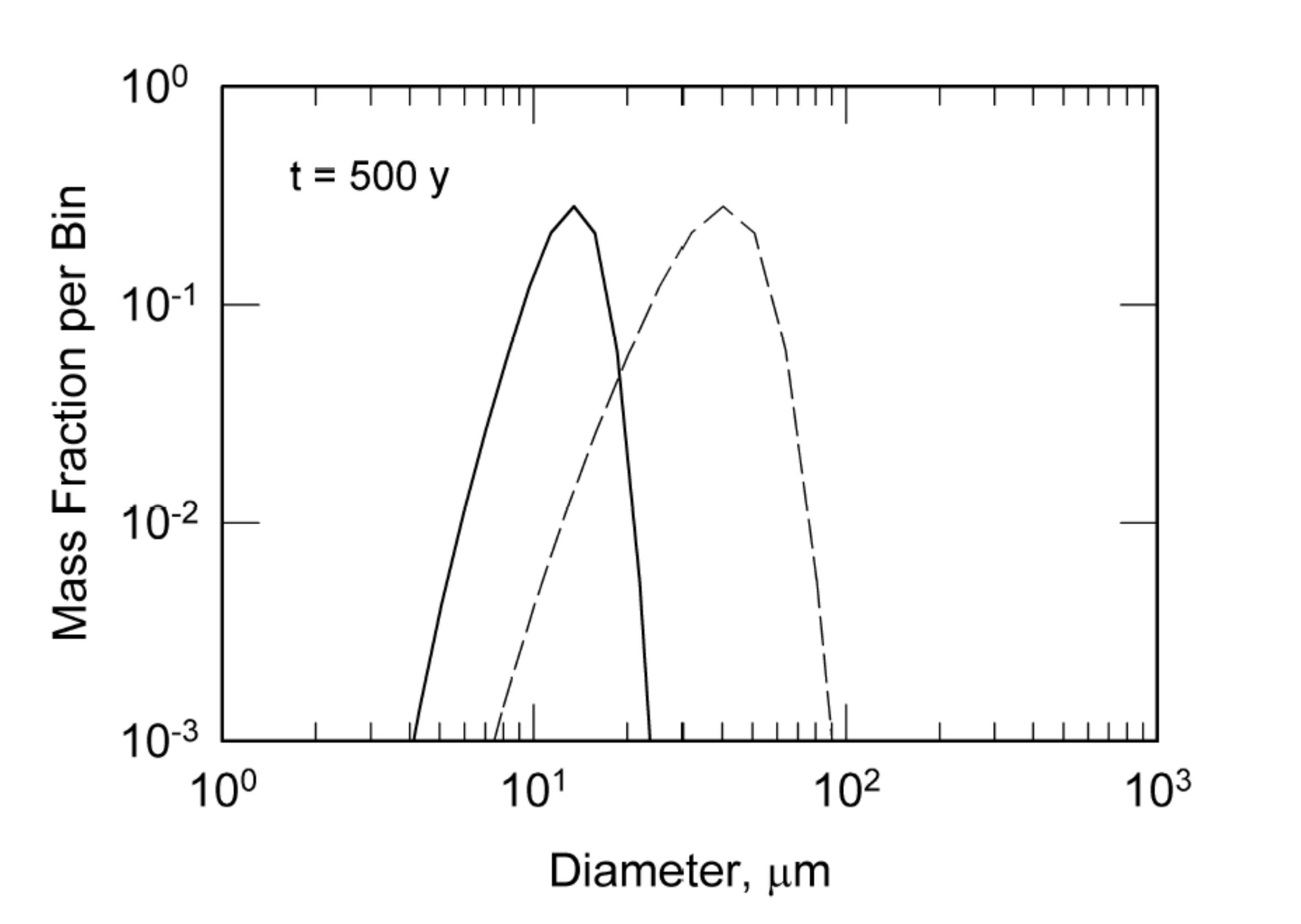}
\includegraphics[width=8.75cm]{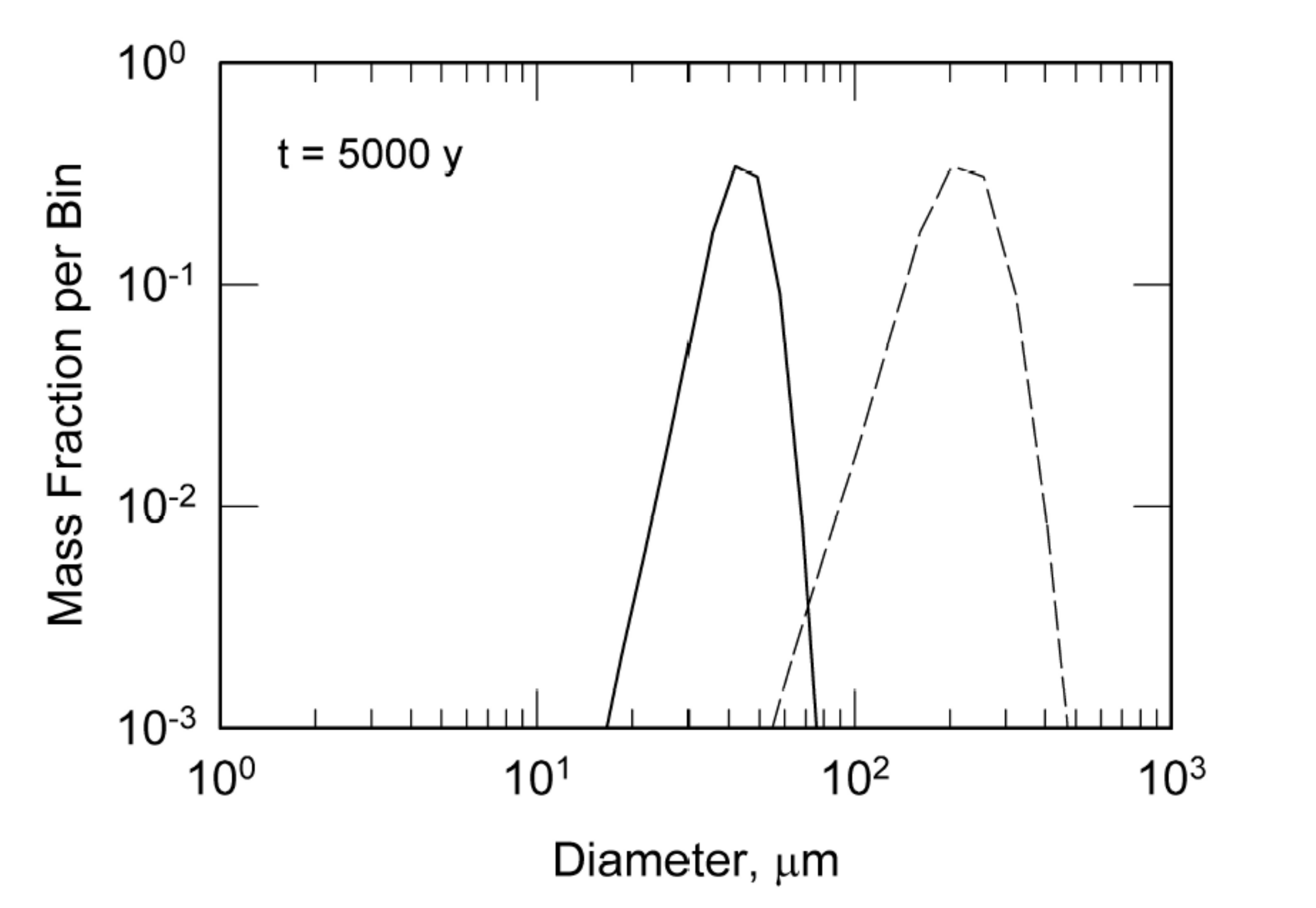}
\caption{Size distributions produced by thermal coagulation, ignoring fragmentation and turbulence.
Nebular conditions are gas density 10$^{-9}$ g cm$^{-3}$, $T$ = 300 K. 
At $t$ = 0, micron-sized grains are present with mass fraction 1.25 x 10$^{-3}$
relative to the gas. Size distributions at 500 years (left) and 5000 years (right)
are shown as mass fractions in logarithmic diameter bins with width 2$^{1/3}$. 
Aggregates are assumed to have fractal dimension 2.1 (dashed lines). Also 
shown are equivalent diameters if fractals are collapsed to compact spheres 
(solid lines).
}
\label{fig:history2}
\end{figure} 

%OLD TEXT SJW

%Although thermally-driven coagulation of small grains can easily yield
%intermediate-sized aggregates, this is not a likely source for the chondrule
%precursors. The latter would be more massive by 3 to 4 orders of magnitude,
%requiring \app 10$^{8}$ monomers, and would take much more time to coagulate under Brownian
%motion. Moreover, purely thermal coagulation yields a narrow unimodal size
%distribution, which would effectively remove the potential source of intermediate-sized
%particles. The chondrule precursors must have another source (or sources),
%such as collisional disruption of larger "pebbles" or planetesimals due to 
%turbulence or differential drift motions induced by nebular gas drag (Weidenschilling
%\& Cuzzi 1993). A collisional source would tend to yield a power-law size
%distribution of larger fragments, but micron-sized fragments or grains released
%by collisions would rapidly coagulate into fractal aggregates. Thus, the overall
%size distribution of solids in the chondrule-forming region would be complex,
%due to superposition of fragmentation and coagulation at larger and smaller sizes.

%REVISED TEXT SJW

The larger chondrule precursors must have some source (or sources) other than purely thermal coagulation. This could include coagulation driven by turbulence or differential drift motions induced by nebular gas drag (Weidenschilling \& Cuzzi 1993), and/or collisional disruption of larger bodies (pebbles, boulders, or planetesimals). A source involving collisional disruption would tend to yield a power-law size distribution of fragments, but micron-sized fragments or grains released by collisions would rapidly coagulate into fractal aggregates. Thus, the overall size distribution of solids in chondrule-forming regions would be due to superposition of simultaneous fragmentation and coagulation, consistent with turbulent coagulation models (Zsom et. al. 2011; Ormel \& Okuzumi 2013; Birnstiel et al. 2014)

\section{Shock Model}

We used the shock code of Morris \& Desch (2010) to carry out our calculations of chondrule thermal histories, with the inclusion of intermediate-sized solids of $a$=10, 15, 20, or 30 $\mu$m.  As in Morris \& Desch (2010), it is assumed that averaged over the nebula, the mass of solids is a fraction $0.005$ of the mass of gas, and that 25\% of the solids' mass is in the form of submicron- to micron-sized dust grains, with the remaining 75\% in the form of large particles (consistent with the proportions in ordinary chondrites).
The fraction of the gas mass that is in large particles is therefore $0.00375$ when averaged over the nebula.
We increase the density of large (chondrule-sized) solids overrun by the shock by a ``concentration"
factor ${\cal C}$ relative to this fraction.  Here, we assume ${\cal C} = 10$, due to settling of solids toward the nebular midplane (Weidenschilling 1980) or turbulent concentration (Cuzzi et al. 2008).  
Consistent with models of the solar nebula (e.g., Desch 2007), we assume an ambient gas density 
$\rho_{\rm g} = 1 \times 10^{-9} \, {\rm cm}^{-3}$.
The speed of the shock is taken to be $8 \, {\rm km} \, {\rm s}^{-1}$. 
This speed, a fraction of the Keplerian velocity $\approx 20 \, {\rm km} \, {\rm s}^{-1}$ at 2 AU, is 
consistent with shocks driven by gravitational instabilities (Boss \& Durisen 2005).

In this study, 1/3 of the ``large" particles are intermediate in size, with radius $a$ = 10, 15, 20, or 30 $\mu$m, and the remaining 2/3 of the large particles have radius, $a$ = 300$\mu$m.  The mass fraction of intermediate-sized particles is chosen to be equal in mass to the micron-sized grains.  This improves on the models of Morris \& Desch (2010) and Desch \& Connolly (2002), which assumed a simple monodispersion of 300$\mu$m precursors mixed with micron-sized dust.  A single, temperature-dependent approximation to the Planck-averaged opacity is used to calculate opacity due to micron-sized dust grains, as in Morris \& Desch (2010).  Opacities for intermediate and large particles are calculated as in Eqn. 4.  Opacities for fractal aggregates are likely strongly enhanced over single grains (Cuzzi et al. 2014), so our pre-shock opacity is a lower limit.  The dust, considered coupled to the gas, evaporates completely and nearly instantaneously at 1500 K (Morris \& Desch 2010).  Chondrules and intermediate-sized solids evaporate at identical, energy-limited rates when their temperatures exceed their liquidus temperature.  Vapor evaporated from the large precursors is added back to the gas.  All other parameters remain unchanged from the shock model of Morris \& Desch (2010).  The intermediate-sized particles and chondrule precursors are assumed to be spherical, with material density 
$3.3 \, {\rm g} \, {\rm cm}^{-3}$.  Strictly speaking, for consistency with the arguments in Section 5, one should start the simulation with only fractal aggregates of micron-sized grains and chondrule precursors, then add the intermediate-sized particles when the small grains reach melting temperature.  However, since the opacity is dominated by the micron-sized grains prior to melting, and the mass in intermediate-sized particles is chosen to be equal to the mass in small grains, our simulation is a good approximation to the conversion of fractal aggregates to compact, intermediate-sized particles.  

The shock model of Morris \& Desch (2010), applicable to large-scale nebular shocks, assumes a 1-D steady-state flow, conserving mass, energy, and momentum.  We summarize the model below and refer the reader to Morris \& Desch (2010) for further details. 

Any shock generated in the solar nebula will heat solids in three ways: 
1) by thermal exchange between the hot, dense, post-shock gas,
2) by frictional heating, as the particles are slowed to the reduced post-shock gas velocity, and
3) by absorption of infrared radiation emitted by heated particles.    
Shocks are transient, non-equilibrium structures, so shock models of chondrule formation must account for the dynamics and energetics of chondrules and 
gas separately, and must include interactions between the two.  Shock models must also account for hydrogen dissociation and recombination, as well as the transfer, absorption and emission of radiation. 
To minimize the complexity of the problem, Morris \& Desch (2010) considered a 1-D geometry, in which physical conditions were
assumed to vary only with distance $x$ from the shock front, with calculations restricted to a range of $x$, the computational domain. 
Use of the 1-D approximation assumes implicitly that the lateral extent of the shock front greatly exceeds the 
computational domain, consistent with large-scale nebular shocks, such as those driven by gravitational instabilities (GI-driven shocks).

The shock code of Morris \& Desch (2010) assumes separate gas and solid fluids. 
The gas is divided into four populations: atomic hydrogen (H), molecular hydrogen (${\rm H}_{2}$), helium 
atoms (He), and molecules resulting from the evaporation of solids, which is represented with SiO.
Solids are divided into three types, submicron- to micron-sized dust grains, intermediate-sized particles of radius $a = 10 \mu$m, and larger, spherical particles of radius $a = 300 \mu$m, representative of chondrule precursors.
Dust grains are assumed to be dynamically and thermally coupled to the gas, sharing its velocity
and temperature.  We employ an approximation to the temperature-dependent mean Planck opacity, described in Morris \& Desch (2010), to calculate the opacity due to dust.  If at any point the dust grains exceed 1500 K, they are assumed to be evaporated from that point forward. 
The intermediate-sized and chondrule-sized particles are unique in their number density $n$, their 
velocity $V$, their temperature $T$, and their radii $a$, as well as material properties.
All fluids are initialized at the pre-shock computational boundary with speed $V_{\rm s}$ and temperature
$T_{\rm pre}$. 
The densities, velocities, and temperatures are integrated forward using a fourth-order Runge-Kutta
integration, assuming a steady-state flow and applying the equations of continuity, a force equation 
including the drag force between gas and solids, and appropriate energy equations. 
The gas can absorb radiation energy (via the dust opacity), can be heated by interaction with chondrules and and their precursors, as well as the intermediate-sized particles, and can lose or gain heat energy through dissociations or recombinations of hydrogen molecules and molecular line radiation of $H_2O$. 
Chondrules, their precursors, and intermediate-sized solids, can emit or absorb radiation, or be heated by thermal exchange with the gas or by frictional 
heating.

Calculation of the radiation field follows the approach outlined in Mihalas (1978), assuming
plane-parallel, temperature-stratified slab atmospheres.
Given the incident radiation fields and the source function, $S(\tau)$ at all optical 
depths, the mean intensity of radiation, $J(\tau)$ (integrated over 
wavelength) is given by:
\begin{equation}
J(\tau) = \frac{I_{\rm pre }}{2} {\rm E}_{2} ( \tau_{\rm m} - \tau )
         +\frac{I_{\rm post}}{2} {\rm E}_{2} ( \tau ) + \frac{1}{2}
         \int_{0}^{\tau_{\rm m}} S(t) {\rm E}_{1} \left| t - 
         \tau \right| \, {\rm d} t,
\end{equation}
where $I_{\rm pre }$ is the intensity of the radiation in the pre-shock region, $I_{\rm post}$ is the intensity in the post-shock region, $\tau$ is the optical depth, and ${\rm E}_{1}$ and ${\rm E}_{2}$ are the exponential integrals.
During the integrations, the radiation field is considered fixed while the gas and particle dynamics 
and energetics are calculated as a function of $x$.
It is assumed that $I_{\rm pre} = \sigma T_{0}^{4}$ in the pre-shock region, and
$I_{\rm post} = \sigma T_{\rm post}^{4}$ in the post-shock region, where $\sigma$ is the Stephan-Boltzmann constant.  
The calculation of $T_{\rm post}$, the post-shock temperature far from the shock front, is
assumed to eventually cool to the background temperature, as described in Morris \& Desch (2010), but see below for caveats.  After all other variables have been integrated across the entire computational domain, the radiation 
field is recalculated based on the updated particle densities, radii and temperatures.
The particles and gas are sent through the shock again, their temperatures calculated based on the 
new radiation field, and the solution is iterated until the radiation field and the particle temperatures are self-consistent.
In this respect, the calculation is a so-called Lambda Iteration, which is known to converge slowly 
(Mihalas 1978). 
Indeed, in our typical runs, convergence of particle temperatures everywhere to within 5 K and 
convergence in $J_{r}$ to within $0.2\%$ were only achieved in several hundred iterations, a 
number comparable to the optical depths across the computational domain.  
A more sophisticated treatment of the radiative transfer is, however, difficult to implement, and is 
subject to many feedbacks.
We chose to use the Lambda Iteration method for its ease of implementation, and for 
its demonstrated stability and convergence in reasonable computing times (hours).

It is important to note, as pointed out in Morris \& Desch (2010) and Stammler \& Dullemond (2014), that large scale nebular shocks are not strictly one dimensional, as radiation will escape vertically from the disk.  Models performed in 1-D will therefore not properly account for vertical energy losses, thereby enabling precise determination of the radiation field and post-shock temperature.  Stammler \& Dullemond (2014) included an energy loss term in their 1-D model in an attempt to account for radiation escaping the top and bottom of the disk, as well as an iterative approximation of post-shock temperature, and found $T_{\rm post}$ remained constant at a higher temperature than the background, as was assumed in Morris \& Desch (2010).  It is clear that a true 2-D model is needed to properly determine the effects of vertical energy loss.
%The reader is referred to Morris \& Desch (2010) for further details of the shock model.   

\section{Results and Discussion}

\begin{deluxetable}{cccc}
\tablecolumns{4}
\small
\tablewidth{0pt}
\tablecaption{Thermal histories of intermediate-sized particles in large-scale nebular shocks.}
\tablehead{
%\multicolumn{11}{r}{{\sc Results from this Study $\;\;\;\;\;\;\;\;\;\;\;\;\;\;\;$  Results from Previous Study}} \\
%\cline{2-12}
\colhead{Radius ($\mu$m)}&
\colhead{$T_{\rm peak}$ (K)}&
\colhead{Cooling rates (K/hr)$^a$}&
\colhead{$t_{\rm pre-heat}$ (min)} 
}
\startdata
10& 1963& 20-40& 16.1 \\
15& 2000& 10-30& 50.6 \\
20& 2000& 10-30& 55.6 \\
30& 2000& 10-30& 74.0 \\
\enddata
\tablenotetext{a}{Cooling rates through crystallization temperature range}
\end{deluxetable}

In this study, we have improved upon the shock model of Morris \& Desch (2010), by considering a simple size distribution of particles, including micron-sized dust, intermediate-sized particles of radius $a$=10, 15, 20 or 30  $\mu$m, and larger chondrule precursors.  A key result is that the intermediate-sized solids present in the pre-shock region significantly enhance the opacity over the case where we only consider larger chondrule precursors.  However, they have largely evaporated in the post-shock region, and therefore do not contribute significantly to the opacity, and therefore the cooling rates of chondrules.  Cooling rates through the majority of the crystallization range are raised only slightly over the canonical case of Morris \& Desch (2010), from \app 10-25 K/hr to 20-40 K/hr, at most.  Cooling rates from 1800 - 1600 K are somewhat higher, but are still well within the range consistent with the meteoritic constraints on thermal histories.  For intermediate-sized particles of radius $a$=15, 20, and 30 $\mu$m, the excess heating in the pre-shock region is reduced, but not eliminated (Table 1).  However, for intermediate-sized particles of radius $a$ = 10 $\mu$m, the time spent at T $>$ 1300 K before the shock is reduced to $<$ 30 minutes (Table 1 and Figure 3), consistent with the constraint on isotopic fractionation of S (Tachibana \& Huss 2005).  In some cases, the newly-formed chondrules may then move into nearby regions of the nebula and mix with existing dust and microchondrules, accreting these components as fine-grained rims.  

% FIGURE 3
\begin{figure}[ht]
\includegraphics[width=14cm]{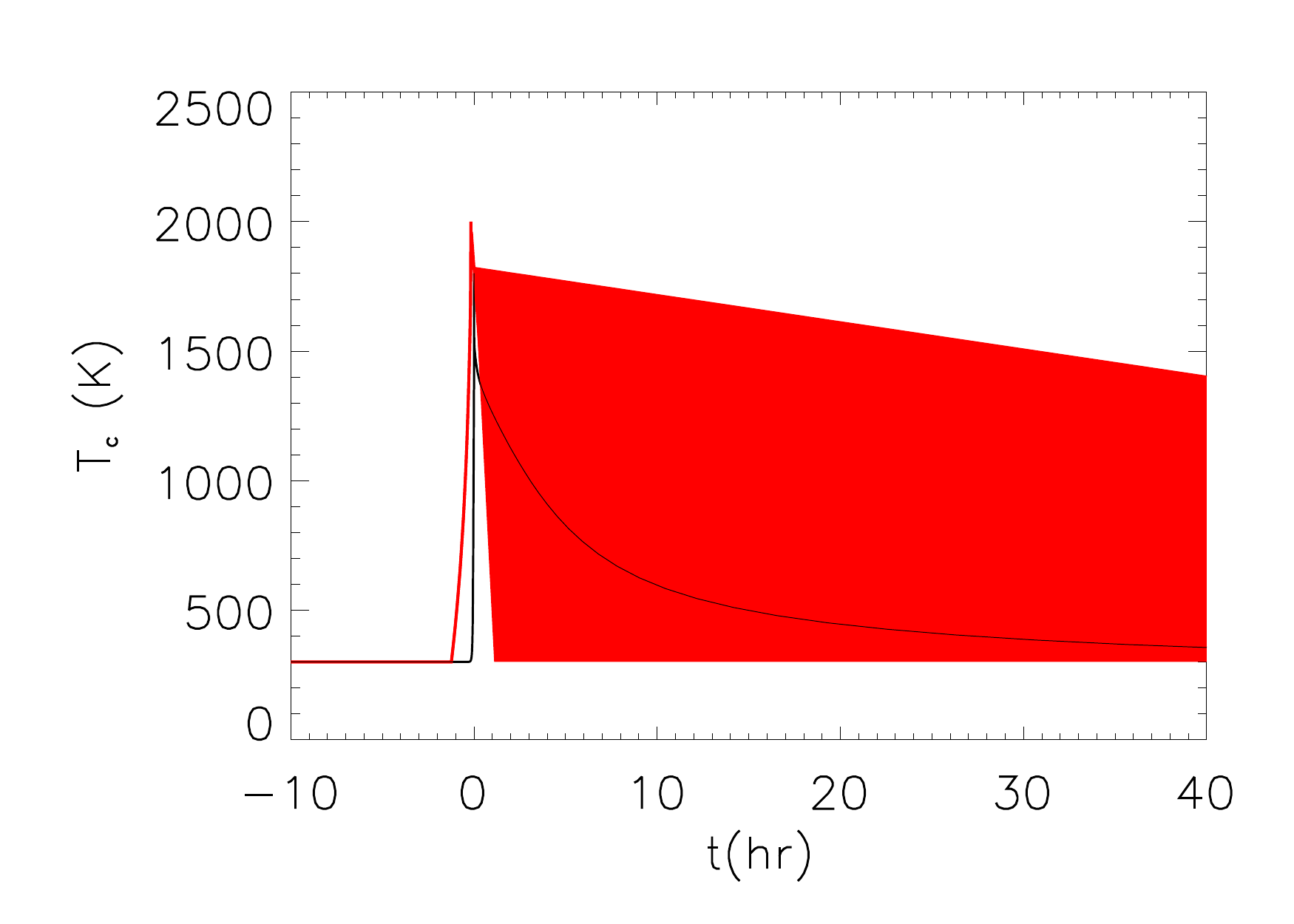}
\caption{The black curve indicates chondrule thermal histories as predicted by the shock model of Morris \& Desch (2010), with the inclusion of ``microchondrules", as compared to those inferred from experimental constraints (red curve).  The shaded region in the post-shock region indicates the range of inferred cooling rates (10-1000 K/hr).  Chondrules start at an ambient temperature of 300K, and remain at \app 300K until \app 15 minutes before the shock front.  }
\label{fig:history2}
\end{figure} 

%\section{Conclusion}

The nebular shock model of Morris \& Desch (2010) met most of the constraints on the thermal histories of chondrules.  However, the model failed to meet the constraint on rapid heating, required to prevent isotopic fractionation of S (Tachibana \& Huss 2005).  In the model, chondrules were heated to temperatures exceeding 1300 K and remained hot for \app 4 hours before reaching their peak temperature.  Our proof-of-concept study here shows that the inclusion of intermediate-sized particles of 10 $\mu$m in radius, eliminates this pre-heating, and post-shock cooling rates of chondrules remain consistent with meteoritic constraints.  Our study has provided the insight that a higher opacity of solids will insulate the pre-shock region, eliminating excess preheating of chondrule precursors.  Particles providing this higher opacity must be large enough to avoid evaporation during the time it takes the shock to traverse the pre-heated region, yet small enough to evaporate in a time short compared to the post-shock chondrule cooling times.  Future experimental studies will determine the melting timescale of fractal aggregates, and future theoretical parameter studies will place limits on size distributions and mass enhancement factors.

 %This model, which is applicable to large-scale shocks only, does not address some of the non-thermal constraints on the chemical environment of chondrule formation. 

%Chondrules appear to have been exposed to anomalously high partial pressures of volatiles. Alexander et al. (2008) have reported high mass fractions of Na in olivine phenocrysts of Semarkona chondrules, strongly implying that the melt contained Na throughout olivine crystallization.  This would require partial pressures of Na vapor high enough to prevent evaporation of Na from the melt.  Potassium in chondrules exhibits a lack of isotopic fractionation (Alexander et al. 2000), implying evaporation of K was likewise suppressed.  FeO contents of chondrules indicate elevated oxygen fugacity during chondrule formation (Krot et al.\ 2000; Fedkin \& Grossman 2006; Grossman et al.\ 2011), requiring water vapor enrichments of \app 240-820 (
%1990; Krot et al. 2000; Fedkin et al. 2011; Grossman et al. 2011).  In such an environment of enriched volatiles, the constraint on rapid heating no longer applies (Connolly et al.\ 2006).  In that case, alternatives to large-scale shocks may need to be explored.  Morris et al. (2012) have suggested that just such an environment existed in bow shocks around planetary embryos.  In their model, an environment rich in volatiles is provided by the outgassed atmosphere of the protoplanet's magma ocean.  Not only does their model match the constraints on thermal histories of chondrules, but it meets the constraints on the chemical environment of chondrule formation as well. 

\acknowledgements

The authors would like to thank J$\ddot{\rm u}$rgen Blum for helpful discussions.  MAM was supported by NASA Origins of Solar Systems Grant NNX10AH34G, NASA Cosmochemistry Grant NNX14AN58G, and NASA Emerging Worlds grant NNX15AH62G.  SJW was supported by NASA Origins of Solar Systems Grant NNX13AI20G.  SJD was supported by NASA Origins of Solar Systems Grant NNX10AH34G.

\end{document}